\documentclass[%
 reprint,
 amsmath,amssymb,
 aps,
 prl,
]{revtex4-2}

\usepackage{graphicx}
\usepackage{dcolumn}
\usepackage{bm}
\usepackage{hyperref}
\usepackage{braket} 
\usepackage{bbold}
\UseRawInputEncoding

\begin{document}

\preprint{APS/123-QED}

\title{Quantum non-demolition measurements of moving target states}

\author{Anton L. Andersen}
\email{anton.andersen@phys.au.dk}
\affiliation{
Center for Complex Quantum Systems, Department of Physics and Astronomy,
Aarhus University, Ny Munkegade 120, DK-8000 Aarhus C, Denmark}

\author{Klaus M{\o}lmer}
\email{moelmer@phys.au.dk}
\affiliation{Aarhus Institute of Advanced Studies, Aarhus University, H{\o}egh-Guldbergs
Gade 6B, DK-8000 Aarhus C, Denmark\\
Center for Complex Quantum Systems, Department of Physics and Astronomy,
Aarhus University, Ny Munkegade 120, DK-8000 Aarhus C, Denmark}

\date{\today}

\begin{abstract}
We present a protocol for probing the state of a quantum system by its resonant coupling and entanglement with a meter system. By continuous measurement of a time evolving meter observable, we infer the evolution of the entangled systems and, ultimately, the state and dynamics of the system of interest. The photon number in a cavity field is thus resolved by simulated monitoring of the time dependent excited state population of a resonantly coupled two-level system, and we propose to regard this as an extension of quantum non-demolition measurements with potential applications in quantum metrology and quantum computing. 
\end{abstract}

\maketitle

{\it Introduction.}
In most studies and applications of quantum systems, it is required to perform precise measurements of a physical observable to either detect its value in a given state or changes of its value due to physical interactions. So-called quantum non-demolition (QND) observations play a special role: these are observations where the interaction with the measurement apparatus does not change the value of the observable of interest or any other property of the system that may subsequently cause changes of that value \cite{Braginsky1980,Caves1980,BraginskyKhalili1990}. 
QND observables permit practical detection schemes where a sequence of weak measurements accumulates measurement statistics and gradually approaches a projective measurement with the outcome distribution given by Born's rule. QND measurements are useful for the high precision monitoring of perturbations or dissipative state changes of quantum systems and sensors \cite{Braginsky1988,BraginskyKhalili1990,Braginsky1996,bocko_measurement_1996}. 

The degree of excitation of a quantum system commutes with the system Hamiltonian and constitutes a QND observable, which can be probed by a dispersive interactions that induce a complex phase rotation on, e.g., a qubit or field probe. Repeated or continuous measurements with such probes gradually project the system of interest on an energy eigenstate \cite{Brune1994,johnson_quantum_2010,lachance-quirion_resolving_2017,NakamuraSciAdv,SAW,BAW,Safavi,Lehnert,lachance-quirion_entanglement-based_2020}, and they may be used to identify, quantum jumps in its excitation dynamics  \cite{gleyzes_quantum_2007,guerlin_progressive_2007,PQSHaroche,Siddiqi,Imamoglu,Reverse,Peil,Schoelkopf2014}. \begin{figure}[t]
\includegraphics[width=\columnwidth]{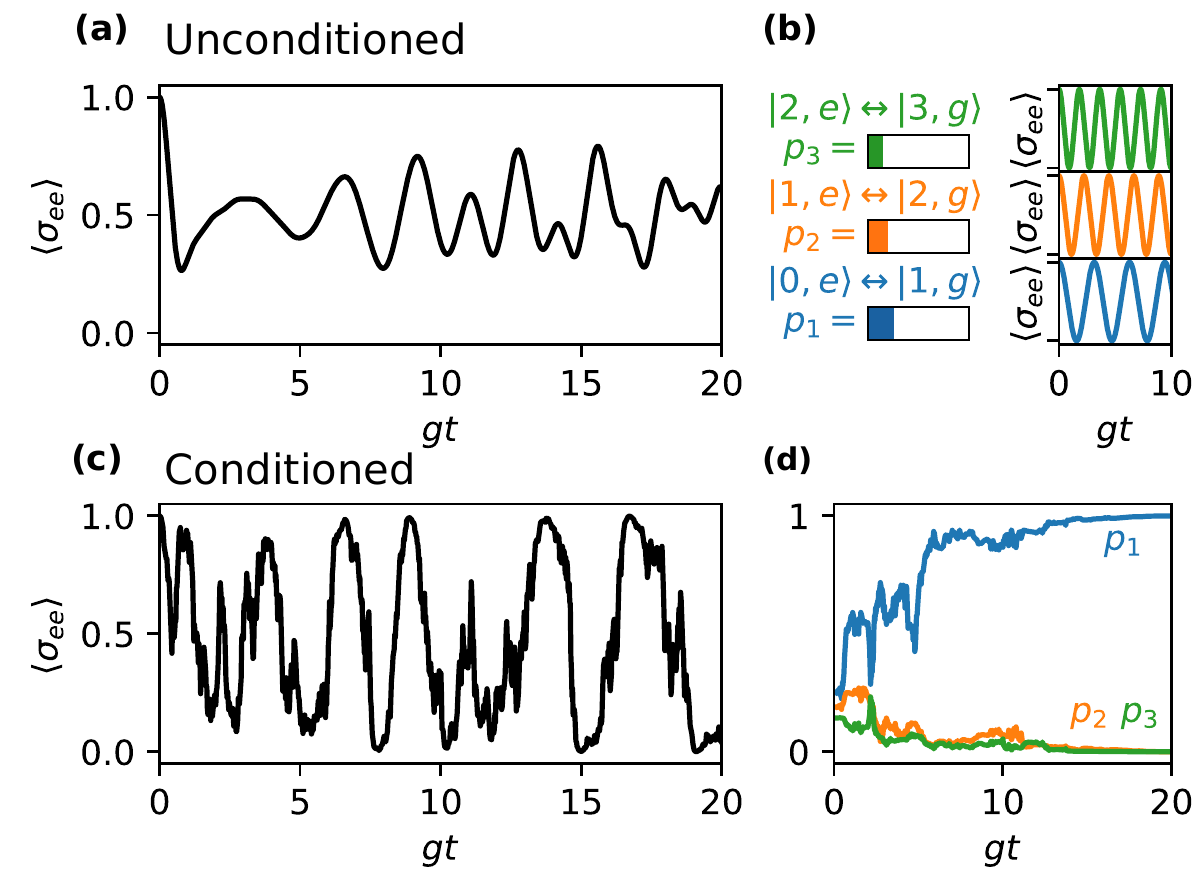}
\caption{\label{fig:1} (a)  Unconditioned dynamics of the excited state population $\langle\hat{\sigma}_{ee}\rangle$, showing collapses and revivals  due to its composition by quantum Rabi oscillations with different frequencies and constant weight factors $p_n$, shown in (b).  (c) The stochastic dynamics of the expectation value $\langle\hat{\sigma}_{ee}\rangle$, conditioned on weak continuous measurement of $\hat{\sigma}_{ee}$. (d) The continuous probing of $\hat{\sigma}_{ee}$ gradually identifies a single Rabi oscillation frequency and hence collapses the system from a thermal ensemble with mean excitation $\langle \hat{n}\rangle=3$ into a single energy subspace as shown by the evolution of the subspace probabilities $p_n$.}
\end{figure}
Variants of QND measurements include stroboscopic QND measurements, such as brief position measurements carried out around times $t_n=n\pi/\omega$ of an harmonic oscillator with frequency $\omega$ \cite{Polzik2011,Polzik2015,Wade2017}, to enable the study of periodically evolving properties of quantum systems, and emergent QND measurements, which probe a physical observable very weakly and effectively, over time, extract its expectation value in one of the energy eigenstates \cite{yang_theory_2018,AndersLuca}. Other strategies employ additional degrees of freedom to evade back action and thereby reach ultimate sensitivity with quantum probes \cite{TsangCaves2010,PolzikEnt2010,TsangCaves2012,Hammerer2014,PolzikHammerer2014}.

{\it Probing by resonant Rabi dynamics.} 
In this Letter we propose a different approach for the measurement of the excitation of an oscillator system, relying on resonant interactions and a deliberate exchange of quanta of energy with a qubit meter system.  Fig.1(a) shows how a mixture or superposition of oscillator eigenstates leads to the so-called damped and revived Rabi oscillations appearing as oscillations in a two-level resonant probe with different $n$-dependent frequencies, demonstrated, e.g., with trapped ions  \cite{meekhof_generation_1996} and superconducting qubits \cite{Martinis}. In this situation, the total, shared number of excitations is a conserved quantity and QND observable, and we suggest to measure its value by a weak continuous monitoring of the oscillating qubit meter excited state population in stead of the final state projective and destructive measurements applied in \cite{meekhof_generation_1996,Martinis}. Fig.1(c) shows the conditioned excited state population dynamics as the measurement gradually resolves the frequency (and phase) of the coherent exchange of energy between the quantum oscillator and the qubit. Panel (d) shows the associated collapse of the system on a moving target state with a definite total number of excitations. We argue that this measurement is faster and may thus enable detection of dynamics and oscilator quantum jumps that cannot be resolved by the dispersive probing. 

{\it Weak continuous measurements.}
We consider a harmonic oscillator resonantly coupled to a qubit with states $\ket{g}$ and $\ket{e}$, via the resonant Jaynes-Cummings Hamiltonian,
\begin{align}
H = \omega (\hat{a}^\dagger \hat{a}+\hat{\sigma}_{ee}) + g(\hat{a}^\dagger\hat{\sigma}_{ge}+\hat{a}\hat{\sigma}_{eg}),    
\end{align}
where $\hbar=1$ such that $\omega$ is the energy spacing of the oscillator and the qubit, $\hat{a}^\dagger$ ($\hat{a}$) is the creation (annihilation) operator of the oscillator, $\hat{\sigma}_{ij}=\ket{i}\!\bra{j}$ and $g$ is the coupling strength. The Jaynes-Cummings coupling will drive oscillations between product states $\ket{n,e}\leftrightarrow\ket{n+1,g}$ with angular frequency $2g\sqrt{n+1}$, as is seen in Fig. \ref{fig:1}. 

We imagine that the qubit is a real or artificial atom with further excited states and that the qubit observable  $\hat{\sigma}_{ee}$ can be continuously measured by phase sensitive, homodyne detection of a classical probe field coupling $\ket{e}$ off-resonantly to an excited state. While this probing is taking place, the dynamics of the system is governed by the stochastic master equation (SME) \cite{howard_m_wiseman_quantum_nodate,blattmann_conditioned_2016},
\begin{align}
d\rho = -i[H,\rho]dt+k\mathcal{D}[\hat{\sigma}_{ee}]\rho dt +\sqrt{2k}\eta\mathcal{H}[\hat{\sigma}_{ee}]\rho dW,
\label{eq:sme}
\end{align}
where  $dW$ represents the Gaussian noise on the phase quadrature of the probe field with mean zero and variance equal to $dt$, $k$ denotes the measurement strength and $\eta$ is the detection efficiency. In this work we will assume $\eta=1$ for simplicity, but our approach works also for non-unit optical detection efficiencies. The first term in Eq. (\ref{eq:sme}) describes the normal time evolution including the Rabi oscillation dynamics. The second term in Eq. (\ref{eq:sme}) contains the dissipation superoperator
\begin{align}
\mathcal{D}[\hat{O}]\rho = 2\hat{O}\rho\hat{O}^\dagger - \{\hat{O}^\dagger\hat{O},\rho\},
\label{eq:D}
\end{align}
describing decoherence due to the disturbance caused by the measurement. The final term in Eq. (\ref{eq:sme}) contains the superoperator
\begin{align}
\mathcal{H}[\hat{O}]\rho = \hat{O}\rho+\rho\hat{O}^\dagger-\langle\hat{O}+\hat{O}^\dagger\rangle_\rho,
\label{eq:H}
\end{align}
where $\langle\hat{O}\rangle_\rho = \text{Tr}[\hat{O}\rho]$. This term describes the back action of the stochastic information gain by the measurement process. The Wiener noise increment $dW$ is given by the difference between the random measurement outcome obtained in the experiment, $dY(t)$, and its expected mean value,
\begin{align}
dY(t) = \langle \hat{\sigma}_{ee} \rangle_\rho dt + \frac{dW}{\sqrt{8k}}.
\label{eq:measurement_record}
\end{align}
$dW$ can be simulated in numerical studies, while one obtains the conditioned dynamics of an experimentally monitored system by solving Eq. (\ref{eq:sme}) with $dW$ extracted from Eq. (\ref{eq:measurement_record}). Numerical solutions to the stochastic master equation are obtained using the QuTiP toolbox \cite{qutip1, qutip2}.\\

\begin{figure*}[t]
\includegraphics[width=\textwidth]{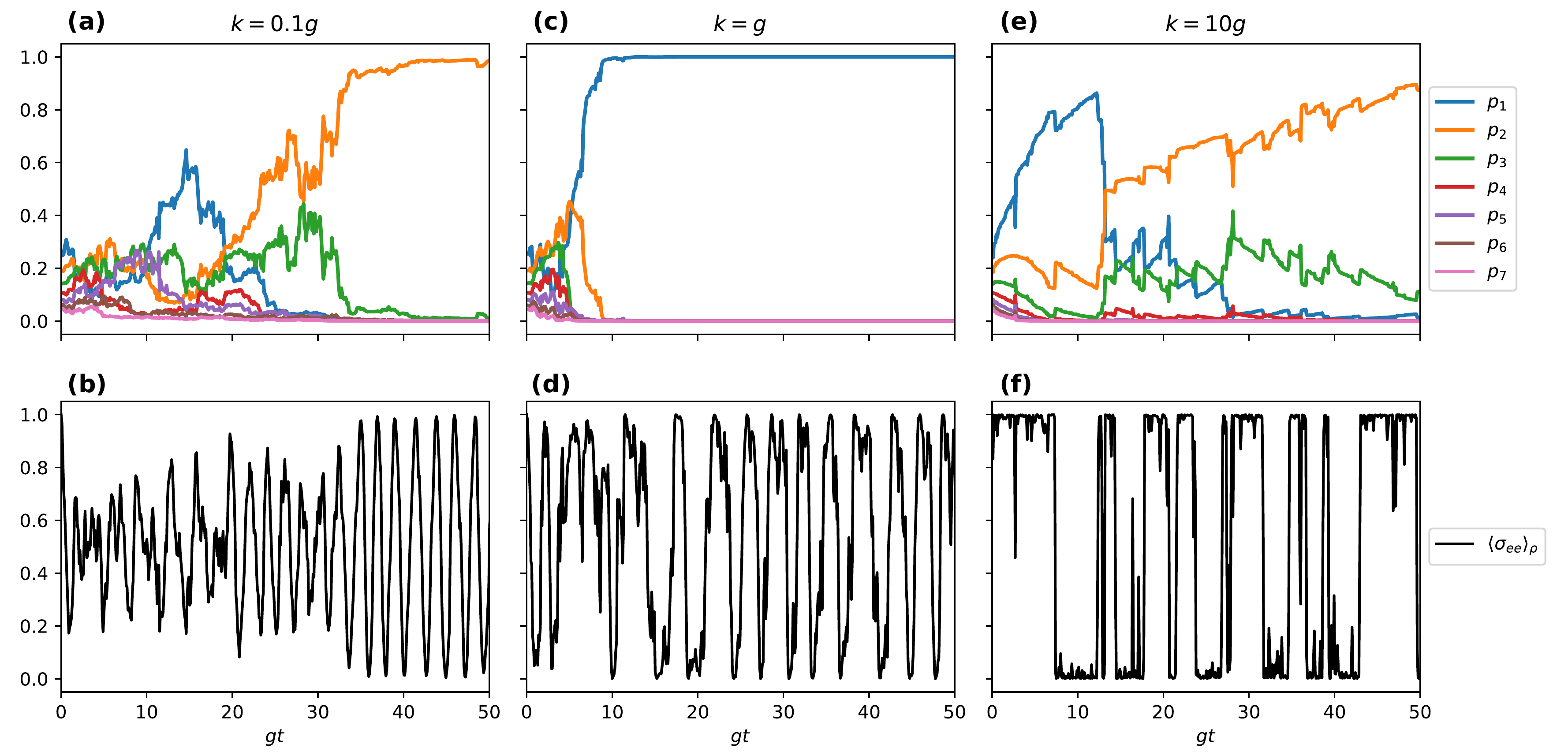}
\caption{\label{fig:regime_plots} Simulated dynamics with the harmonic oscillator prepared initially in a thermal state with mean excitation number $\langle \hat{n}\rangle=3$ and the qubit prepared in the excited state. The upper panels show the probabilities for the total (integer) number of excitations, while the lower panels show the qubit excited state population for weak probing ($k=0.1g$, panels (a) and (b)), strong probing ($k=g$, panels (c) and (d), and very  strong probing ($k=10g$, panels (e) and (f)).}
\end{figure*}

To observe how the continuous measurement of $\hat{\sigma}_{ee}$ reveals the oscillator dynamics, we will consider a situation where the qubit is initially prepared in the state $\ket{e}$, while the harmonic oscillator is in a mixed state described by $\rho_{HO}=\sum_n p_n(t=0)\ket{n}\!\bra{n}$. Results of simulations are presented in Fig. \ref{fig:regime_plots} for $\eta=1$, and they show that the system converges to states with a definite total number of excitations. For the case of weak probing we see that the definite value of $n$ occurs together with a definite harmonic evolution of the excited state population (at frequency $2g\sqrt{n+1}$), while intermediate probing strengths $k$ also identify $n$ but continuously disturb the phase of the Rabi oscillations. For even stronger probing, a Zeno effect prevents the coherent Rabi oscillations and makes the distinction harder between different values of $n$ \cite{Zeno}. 

\begin{figure}
\includegraphics[width=\columnwidth]{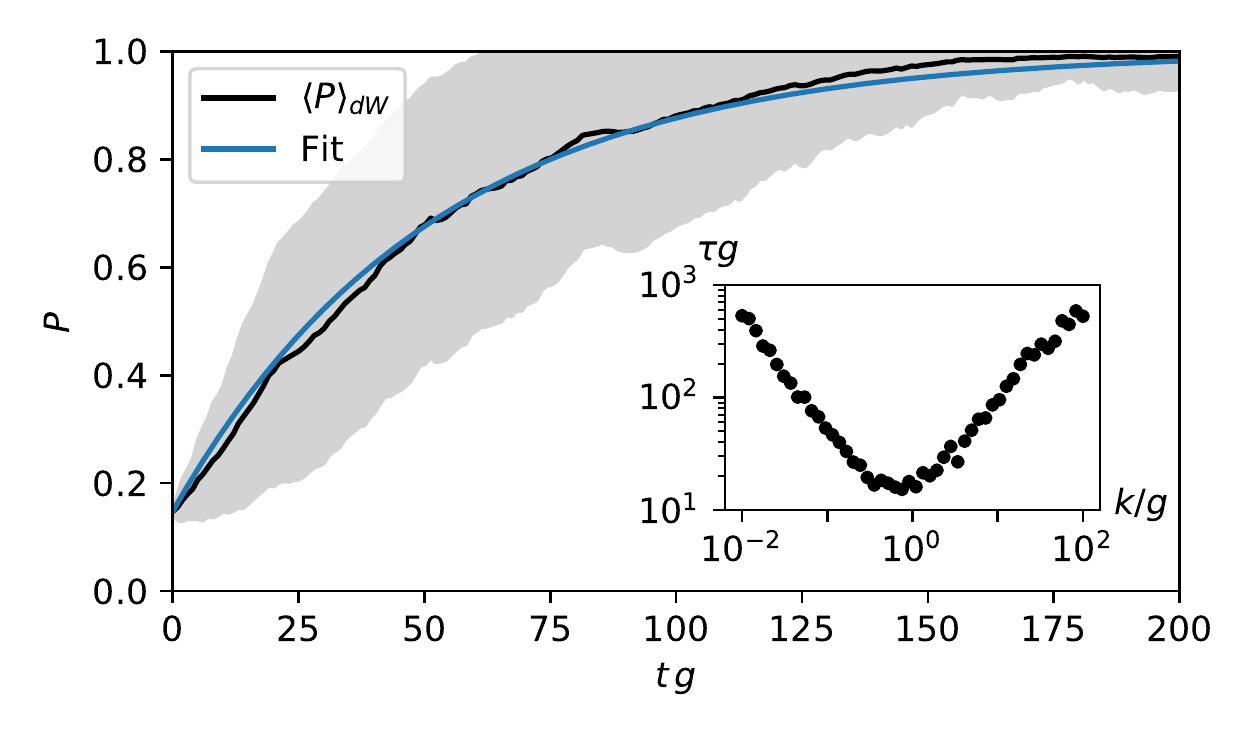}
\caption{\label{fig:WR_avg_purity} The average purity $\langle P \rangle_{dW}$ from 200 simulated trajectories in the weak probing regime (k=0.1g). The gray shaded area corresponds to values within one standard deviation from the mean. The average time  $\tau$, needed to perform the measurement of the energy, is extracted by fitting the model $P(t)=1-(1-P(t=0))e^{-t/\tau}$. The harmonic oscillator is prepared with an initial thermal distribution with $\langle \hat{n}\rangle=3$ and the qubit is initially prepared in the excited state. The insert shows the average time of purification and distinction of the excitation of the harmonic oscillator as function of probing the strength $k$.}
\end{figure}

{\it Optimal probing.} 
In order to assess the time needed for the continuous measurements to determine the degree of excitation of the system by the corresponding frequency of the Rabi oscillations, we study the convergence towards unity of the purity $P$ of the conditional density matrix Tr$(\rho^2)$. As seen in Fig. \ref{fig:WR_avg_purity}, we can fit its mean value over many trajectories with the model $P(t)=1-(1-P(t=0))e^{-t/\tau}$, and repeating this procedure for different probing strengths we observe in the insert of Fig. \ref{fig:WR_avg_purity} that the time needed to perform the QND measurement is smallest in the intermediate strength probing regime, $k\simeq g$. 

This result can be explained qualitatively, since  probing with a small value of $k$ only yields an appreciable signal-to-noise  when accumulated over times $\propto 1/k$.  During that time the system undergoes one or several Rabi oscillations, and despite the white noise component in the weak probe signal it is possible to discern a single leading harmonic component and hence reveal the value of $n$.

While increasing $k$ increases the data extraction rate, when $k$ becomes of the order of the value of $g$ the back action of the qubit excited state measurements causes significant disturbance of the Rabi oscillations. Discerning different $n$-values by the frequency of the Rabi oscillations, is gradually hampered by these disturbances when $k$ exceeds $g$. Ultimately, when $k$ is very large, the measurements effectively project the qubit in its energy eigenbasis and thus freezes the Rabi oscillations by the Quantum Zeno mechanism \cite{Zeno}, see Fig. \ref{fig:regime_plots}(f). 

We note that the moderate and strong measurement back action do not invalidate the QND property with respect to distinction of Rabi subspaces, they only cause stochastic modifications of the harmonic population oscillation within the subspaces  as shown in Fig.3, and hence they make the distinction between different subspaces less effective.

{\it Observation of Quantum Jumps}.
Our probing may be applied to mechanical oscillators, quantized fields, photons and magnons, which are all systems where there has been an interest in demonstrating the quantized nature of their interactions \cite{guerlin_progressive_2007,johnson_quantum_2010,NakamuraSciAdv,SAW,BAW,Safavi,Lehnert,lachance-quirion_resolving_2017, oconnell_quantum_2010,lachance-quirion_entanglement-based_2020,AndersLuca}, and dynamical features such as quantum jumps \cite{gleyzes_quantum_2007,Siddiqi,Imamoglu,Reverse,Peil,Schoelkopf2014}. The latter experiments are often hampered by the time between jumps being comparable to the time needed to detect the change of $n$ in an experiment. For this, our scheme may be particularly useful, and we now discuss how to incorporate thermal quantum jumps in the formalism and how well they are inferred from a measurement record.           

If the  the harmonic oscillator is connected to a thermal reservoir with an average number of excitations $n_T$ and coupling rate where $\gamma$, Eq. (\ref{eq:sme}) is modified into
\begin{eqnarray}
d\rho = &-i[H,\rho]dt-\frac{\gamma}{2}(n_T+1)\mathcal{D}[\hat{a}]\rho dt - \frac{\gamma}{2}n_T\mathcal{D}[\hat{a}^\dagger]\rho dt
\nonumber \\
& +k\mathcal{D}[\hat{\sigma}_{ee}]\rho dt +\sqrt{2k}\eta\mathcal{H}[\hat{\sigma}_{ee}]\rho dW. 
\end{eqnarray}
The terms involving $\mathcal{D}[\hat{a}]$ and $\mathcal{D}[\hat{a}^\dagger]$ are responsible for the loss or absorption of excitations to or from the bath. \\

\begin{figure}
\includegraphics[width=\columnwidth]{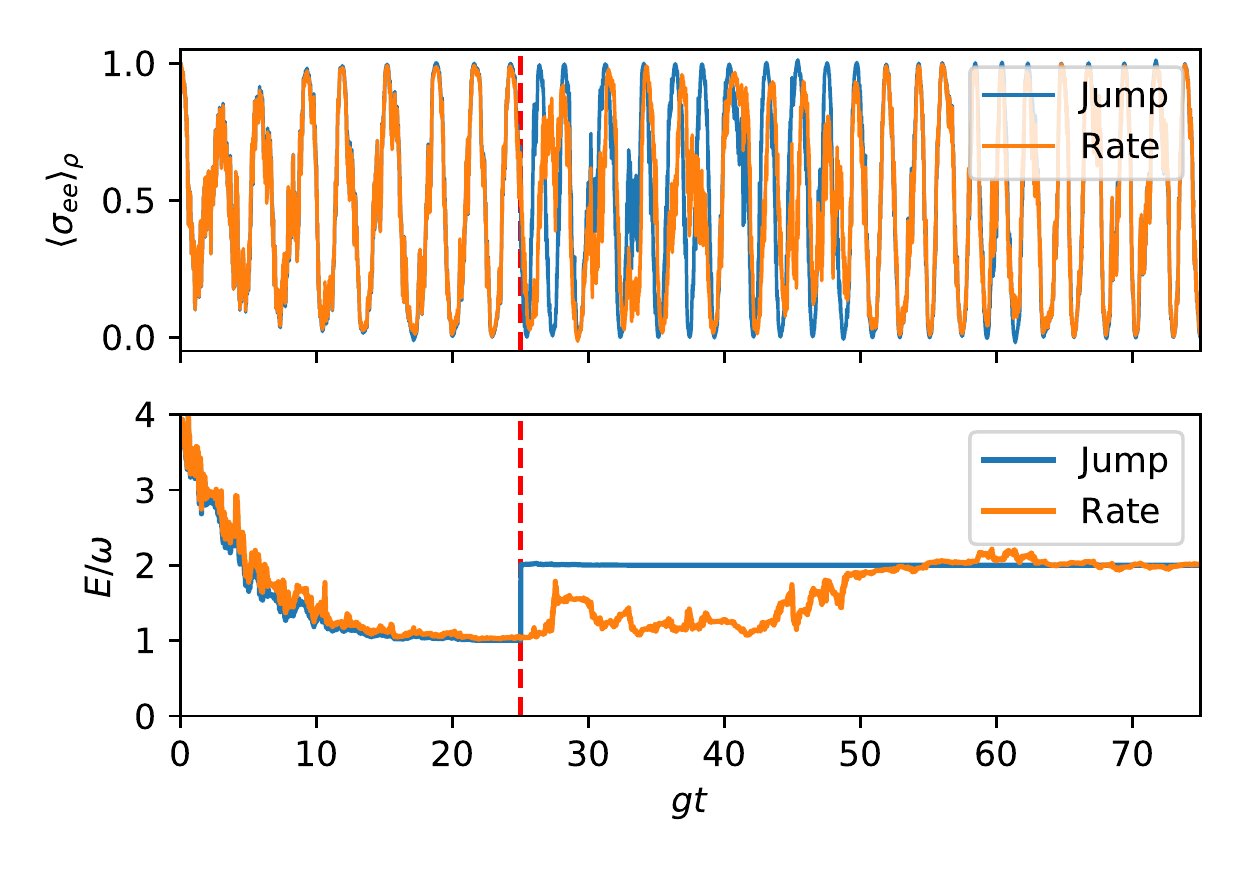}
\caption{\label{fig:jumps_plots} Observation of a quantum jump. The orange curve in the upper (lower) panel shows the  excited state population (average excitation of the oscillator) inferred from weak continuous measurements on the qubit meter. The oscillator is subject to a single quantum jump occurring at $gt=25$, and the blue curves show the inferred qubit excited state population and oscillator number of quanta, assuming the added knowledge of when the jump happened. Parameters used for the simulation are $k=0.1g$, $\gamma=10^{-3}g$, $\langle\hat{n}\rangle=n_T=3$.}
\end{figure}

Figure \ref{fig:jumps_plots}. shows a simulation of the dynamics described by Eq. (6). For this particular simulation, we assumed that no quanta were emitted into or absorbed from the bath until $gt=25$ where we simulated an incoherent heating event. The blue curve in the lower panel shows the mean excitation of the oscillator, inferred by a hypothetical observer of both the probing dynamics and the occurrence of the energy exchange between the oscillator and the heat bath, while the orange curve shows the mean excitation of the oscillator inferred by an observer having only access to the continuous probing record.  In the upper panel, the change in $n$ accompanies a change in the Rabi oscillation frequency appearing instantly in the regular blue curve inferred by the hypothetical observer. The more erratic orange curve reveals the uncertainty of the real observer who realizes the change of state and agrees with the hypothetical observer only after the signal-to-noise ratio has accumulated to permit distinction of the different $n$-values.       

\begin{figure}
\includegraphics[width=\columnwidth]{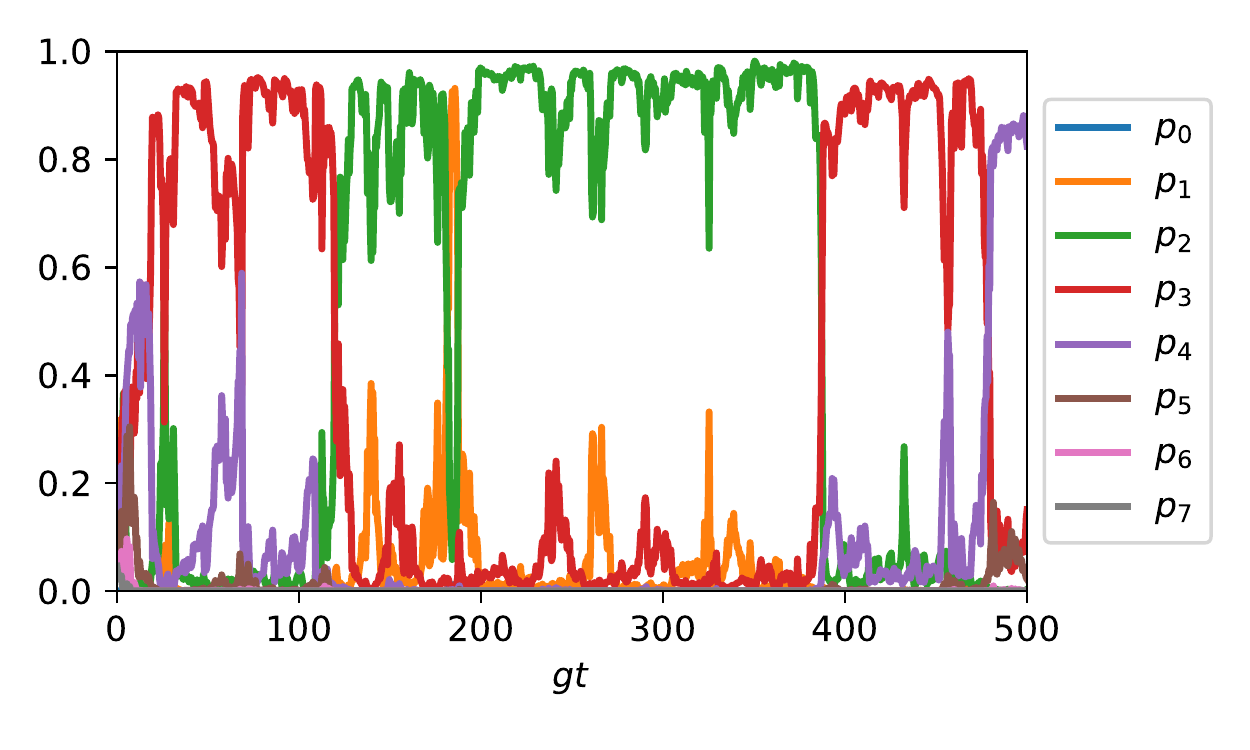}
\caption{\label{fig:many_jumps} Observation of multiple quantum jumps. The harmonic oscillator starts in a thermal distribution with mean excitation number $\langle\hat{n}\rangle=3$. The probing strength $k=g$, such that it is close to the optimal value. The coupling to the bath is $\gamma=10^{-3}g$ and its mean excitation is $n_{T}=3$.}
\end{figure}

In Fig. \ref{fig:many_jumps} we show a longer measurement record with multiple jumps inferred as the rapid transfer of near unit probability weights on different values of $n$. A long time average of these probabilities would reveal the Boltzmann distribution, while the measurements act like a Maxwell demon and turn the probabilities into random almost certain knowledge about the state of the system. We note that, as in \cite{gammelmark_past_2013,PQSHaroche}, it is possible to use the entire measurement record and not only previous data for the theoretical estimation of the state at any given time and that would improve the agreement between the inferred and the true jumps in Figs. \ref{fig:jumps_plots} and \ref{fig:many_jumps}.

{\it General picture.} The characteristic property of our emergent subspace QND procedure is the convergence and subsequent restriction of the system to follow trajectories within single degenerate subspaces of a certain operator $\hat{A}$ which commutes with the Hamiltonian, $[\hat{A},\hat{H}]=0$.  An interaction term in the Hamiltonian $\hat{H}$ causes a time evolution of a meter observable $\hat{B}$, which commutes with $\hat{A}$, and the temporal outcome of measurements of $\hat{B}$ may gradually collapse the system on a state that is evolving in a definite eigenspace of $\hat{A}$. For this detection to work, it is important that the characteristic measurement records differ when the system occupies different such subspaces. In our example, $\hat{A}$ is the total number of excitations and $\hat{B}$ is the qubit meter excitation, and the Rabi oscillation frequencies, indeed, have distinct values in each eigenspace of $\hat{A}$.  Notably, the measurements both reveal the subspace (of $\hat{A}$) and the actual time dependent entangled state of the system and meter from which we infer the separate system dynamics. While our analysis used the example of system and meter entangled state dynamics, the commutator requirements between $\hat{H}$ and $\hat{A}$ and between $\hat{A}$ and $\hat{B}$ suffice for our scheme to resolve the value of $\hat{A}$ by the measurement of $\hat{B}$ in any quantum system, e.g., pertaining to the population of subspaces of a single multi-level quantum system.   

If we assume unit detector efficiency, and an initial mixture of pure states $\ket{\psi_n(t)}$, each occupying a single degenerate subspace of $\hat{A}$,
\begin{align}
\rho(t) = \sum_n p_n(t)\ket{\psi_n(t)}\!\bra{\psi_n(t)},
\label{eq:rho}
\end{align}
we may generalize \eqref{eq:sme} to the measured observable $\hat{B}$ 
\begin{align}
d\rho = -i[H,\rho]dt+k\mathcal{D}[\hat{B}]\rho dt +\sqrt{2k}\eta\mathcal{H}[\hat{B}]\rho dW.
\label{eq:smegen}
\end{align}

The ansatz in  Eq.(\ref{eq:rho}) then leads to the following equations
\begin{align}
dp_n = \sqrt{8k}p_n\Big(\bra{\psi_n(t)}\hat{B}\ket{\psi_n(t)}-\langle\hat{B}\rangle_{\rho(t)}\Big)dW,
\label{eq:dp_n}
\end{align}
and we observe that, if only one state $\ket{\psi_n}$ is populated, $p_n=1$, $\bra{\psi_n}\hat{B}\ket{\psi_n} = \langle\hat{B}\rangle_\rho$, and the stochastic noise terms does not affect the future evolution of the unit value of  $p_n(t)$, while the state $\ket{\psi_n(t)}$ may still evolve within the given occupied subspace. 

To further understand why the system collapses on a single subspace, we note that the purity of the system is $P(t)=\sum_n p_n^2(t)$, and  applying Itô's rule for $d(p_n^2)$ yields
\begin{align} 
\begin{split}
dP = \sum_n \Big[&8kp_n^2\Big(\bra{\psi_n}\hat{B}\ket{\psi_n}-\langle\hat{B}\rangle_\rho\Big)^2dt \\
&+ \sqrt{32k}p_n^2\Big(\bra{\psi_n}\hat{B}\ket{\psi_n}-\langle\hat{B}\rangle_\rho\Big)dW\Big].
\end{split}
\end{align}
The average of $dW$ is zero and hence the average evolution of the purity obeys
\begin{align}
d\langle P \rangle_{dW} = \sum_n 8k\left\langle p_n^2\Big(\langle\hat{B}\rangle_n-\langle\hat{B}\rangle_\rho\Big)^2 \right\rangle_{dW} dt,
\label{eq:dP/dt}
\end{align}
which is positive and causes $\langle P \rangle_{dW}$ to increase until the time evolution of $\langle\hat{B}\rangle_\rho$ is indistinguishable from the one in just one of the subspaces $\langle\hat{B}\rangle_n$. If several subspaces display the same evolution, they are not distinguished and our measurement is emergent QND with respect to their union, but we may populate a mixed state in that union.

{\it Conclusion and Outlook.}
We have presented a new principle for continuous QND measurements which does not project the system on the eigenstate of the QND observable but rather on a still evolving state within a subspace of states. These subspaces are discerned by the characteristic frequency of the evolution of the mean value of the observed quantity, which may be monitored faster than the accumulation of dispersive phase shifts in the more usual QND setting. If $\Delta$ is the detuning and $g$ is the resonant coupling strength between a harmonic oscillator and a two-level system, their dispersive coupling is given by $g^2/\Delta$ \cite{QOBook}. To avoid transfer of excitation, the detuning should be much larger than the coupling strength $g\ll\Delta$, and the timescale on which the conventional QND measurement takes place is longer by a factor of order $\Delta/g \gg 1$ compared to our resonant proposal. 
Our method may pave the way to monitor thermal quantum jumps in real time and use measurements and feedback for rapid state preparation and control.

{\it Acknowledgment.} The authors acknowledge support from the Danish National Research Foundation through the Center of Excellence for Complex Quantum Systems (Grant agreement No. DNRF156). KM acknowledges discussions with Dr. Faezeh Pirmoradian in the early stages of the project.

\bibliography{manuscript}

\end{document}